\documentclass{articlek}

\renewcommand{\refname}{REFERENCES}
\def\be{\begin{equation}}
\def\ee{\end{equation}}

\def\BibTeX{{\rm B\kern-.05em{\sc i\kern-.025em b}\kern-.08em
            T\kern-.1667em\lower.7ex\hbox{E}\kern-.125emX}}

\usepackage{graphicx}
\begin{document}
\sloppy
\twocolumn[{
\vspace*{1.7cm}
\begin{center}
{\large\bf PHASE DIAGRAM OF THE SPIN-1 ANISOTROPIC HEISENBERG MODEL   
WITH A SINGLE-ION ANISOTROPY}\\

{\small J. Dely, jan.dely@upjs.sk, J. Stre\v{c}ka, jozef.strecka@upjs.sk and L.
\v{C}anov\'a, lucia.canova@upjs.sk}\\
{\small Department of Theoretical Physics and Astrophysics, Faculty of Science, P. J.
\v{S}af\'arik University, \\ Park Angelinum 9, 040 01 Ko\v{s}ice, Slovak Republic}

\end{center}
\vspace*{1ex}

{\bf ABSTRACT.} The spin-$1$ anisotropic Heisenberg model with a single-ion anisotropy
is studied using the Oguchi's pair approximation. Although the theory is developed for
lattices with general coordination number, we treat in detail the three-dimensional
lattice with the lowest coordination number, i.e. diamond lattice, where the critical
and tricritical behavior of the system is analyzed as a function of both the
single-ion anisotropy and exchange anisotropy.\\    
}]
\section{INTRODUCTION}

The investigation of critical behavior of quantum Heisenberg systems, which are useful
for the description of many magnetic materials, belongs among the most plentiful areas
of statistical physics. In recent years, the special attention has been devoted to the
low dimensional spin-1 Heisenberg models $[1$-$4]$. Pan and Wang analyzed the effect
of a single-ion anisotropy on the ground-state properties of the spin-1 Heisenberg
model for a general lattice \cite{PA_1}. However, as far as we know, similar treatment
of thermodynamic properties at non-zero temperatures has not been done up to now.
Therefore, the purpose of this paper is to study the criticality and tricriticality of
the spin-1 anisotropic Heisenberg model with a single-ion anisotropy for lattices with
general coordination number.    

\section{MODEL AND ITS SOLUTION}

The Hamiltonian of the spin-1 Heisenberg system in a presence of the single-ion
anisotropy $D$  and the external magnetic field $h$ is described as 
\begin{eqnarray}
H &=& -\sum_{(i,j)} J[\Delta (S_i^xS_j^x + S_i^yS_j^y) + S_i^zS_j^z] \nonumber \\&&
         - D\sum_{i} (S_i^z)^2 - h\sum_i S_i^z, 
\end{eqnarray}
where the first term represents the anisotropic exchange interaction. $J$ is the
exchange coupling constant $(J>0)$ restricted to the nearest-neighbour pairs of spins
and $\Delta$ is the exchange anisotropy parameter ($\Delta$=0 and $\Delta$=1 correspond
to the Ising and isotropic Heisenberg models, respectively). Finally, $S_i^{\gamma}$
($\gamma = x,y,z$) are the components of the spin-1 operator at sites $i$.  

Since the inplementation of single-ion anisotropy $D$ in the system under
investigation can potentially lead to first-order phase transitions, we need to
know the expression of free energy in order to distinguish the stable and unstable
magnetic phases. Due to this fact we adopt the simple Oguchi's pair approximation (OA)
\cite{OG_1}, which is superior to the mean-field method. Thus, the Hamiltonian (1) in
the OA for a cluster with two spins is given by  
\begin{eqnarray}
H_{ij}^{OA}&=&- J[\Delta (S_i^xS_j^x + S_i^yS_j^y) + S_i^zS_j^z] \nonumber \\&&
         - D[(S_i^z)^2 + (S_j^z)^2] - h_{ef} (S_i^z + S_j^z), 
\end{eqnarray}
in which $h_{ef}$ is the effective field, acting on both the spins of cluster, of the
form $h_{ef} = J(q-1)m + h$. $q$~ is~ the coordination number of the lattice and $m$ is the
magnetization per site, i.e. $m = \langle \frac{1}{2} (S_i^z + S_j^z)\rangle$.

The Hamiltonian $H_{ij}^{OA}$ in the matrix representation calculated within a
standard basis set of functions $| S_i^z,S_j^z \rangle$ ($S_i^z = \pm 1,0$ and $S_j^z
= \pm 1,0$) can be written in the form of $9 \times 9$ matrix. Then it is quite
straightforward to obtain nine eigenvalues by diagonalizing the matrix of
$H_{ij}^{OA}$. Thus, the partition function $Z = \textrm{Tr}_{ij}$exp$(-\beta H_{ij}^{OA})$ has
the following form 
\begin{eqnarray}
Z&=&e^{ \beta (-J + 2D)} + 2e^{\beta (J+2D)} \textrm{cosh} ( 2\beta h_{ef} ) \nonumber \\&& 
+ 2e^{\beta (D - J/2)} \textrm{cosh} [\beta \sqrt {(D - J/2)^2 + 2J^2
\Delta ^2}] \nonumber \\&&
+ 4e^{\beta D} \textrm{cosh} (\beta J \Delta) \textrm{cosh} (\beta h_{ef} ), 
\end{eqnarray}
where $\beta = 1/k_BT$ is the reciprocal temperature ($k_B$ is the Boltzmann
constant). The knowledge of the partition function allows us to express the relations for
all thermodymanic quantities. The free energy of the two-spin cluster is then defined as  
\begin{eqnarray}
f = -\beta ^{-1} \textrm{ln} Z + J(q-1)m^2,
\end{eqnarray}
wheares the magnetization $m$ per site is obtained by minimizing the free energy
$(4)$ and is given by  
\begin{eqnarray}
m &=& \frac{1}{Z} \{ \vspace{-0.5cm}  2e^{ \beta (J + 2D)}\textrm{sinh} (2 \beta h_{ef}) 
\nonumber \\&&  \quad + 2e^{\beta D} 
\textrm{cosh} (\beta J \Delta) \textrm{sinh} (\beta h_{ef})\}.
\end{eqnarray} 
Through the use of Eqs. $(4)$ and $(5)$ we can completely analyze the phase diagrams
of the present model.

Indeed, close to the second-order phase transition from the ordered phase ($m \neq 0$)
to the paramagnetic one ($m = 0$) the magnetization $m$ is very small, for $h = 0$,
and one may expand Eq. (5) into the form  
\begin{eqnarray}
m = am + bm^3 + cm^5 + ... ,
\end{eqnarray}
where the coefficients $a,b$, and $c$ are functions of $T, J, D, \Delta$, and $q$. So,
hereby the second-order transition line occurs when $a = 1$, $b < 0$ and the tricritical
point is determined by conditions $a = 1$, $b = 0$, and $c < 0$. On the other hand, the
first-order phase transitions between different phases can be obtain from a comparison
of their free energies.   

\section{RESULTS AND DISCUSSION}

Now, let us study the phase diagram of the system and the thermal 
variations of magnetization for the case of diamond lattice ($q = 4$) in zero external
magnetic field. The choice of this lattice is motivated by the fact that the
diamond lattice is a three-dimensional lattice with the lowest coordination number and
hence, one should expect the most obvious impact of quantum fluctuations on the
cooperative ordering.

\begin{figure} [h,t]                    
\begin{center}                        
\includegraphics[width=80mm]{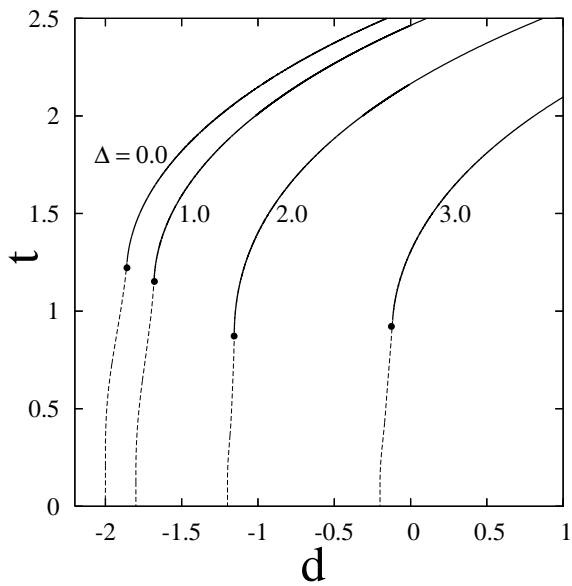}   
\end{center}                         
\vspace{-6mm} \caption{The phase diagram of the spin-$1$ anisotropic Heisenberg model.}
\vspace*{-0.27cm}
\end{figure}                       

The phase diagram in the reduced units $d=D/J$ and $t=k_BT/J$
for different values of $\Delta$ is shown in Fig.~$1$. The solid and dashed lines
represent the second- and first-order phase transitions, respectively. The black circles
denote the positions of tricritical points 
$(TCP)$. As one can see from this figure, the temperature of second-order phase
transitions gradually declines with decrease of $d$, for any value of $\Delta$, until the $TCP$ is 
reached, where the phase transitions change from the second-order to first-order ones. By
further decreasing of $d$ the correspondent first-order phase transition lines fall
smoothly to zero. It is noteworthy that the temperature of second-order phase transitions,
in the limit $d \rightarrow \infty$, does not depend on the anisotropy parameter $\Delta$
and is equal to $t^{\star} = 4.5392$. Hereafter, it is evident from Fig. $1$ that the
present system exhibits the highest critical temperatures in the Ising limit $(\Delta =
0)$. On the other hand, the exchange anisotropy strengthening gradually decreases the 
critical temperature as a result of raising quantum fluctuations. Finally, an increase of
the anisotropy parameter $\Delta$ causes a shift of $d$-coordinate of tricritical point
($d_t$) to the higher values.  
To illustrate the effect of single-ion anisotropy on the phase transitions we show in Fig.~$2$ the
thermal variations of magnetization $m$ in the case of isotropic model $(\Delta = 1)$ for some
appropriate values of $d$. It is obvious from the figure that for $d = -1.65$ and $-1.68$
the magnetization falls smoothly to zero when temperature approaches the critical temperature
$t_c$. This behavior of magnetization is typical for the continuous phase transitions and  
persists until $d > d_t$ ($d_t = -1.6801$ for $\Delta = 1$). 
\begin{figure} [h,t]                    
\begin{center}                        
\includegraphics[width=80mm]{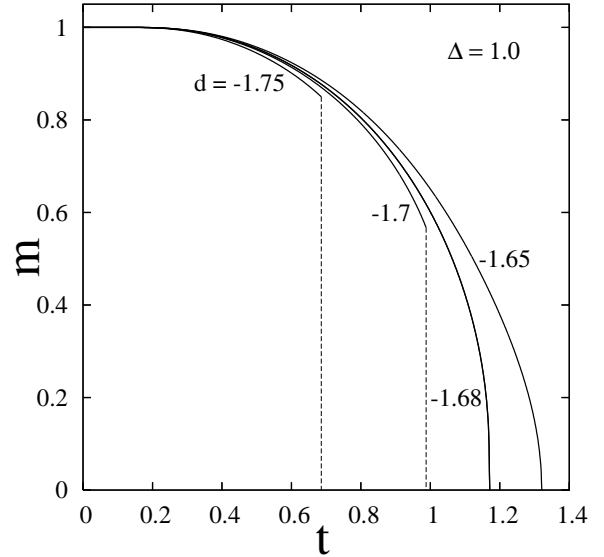}   
\end{center}                         
\vspace{-4mm} \caption{The temperature behavior of magnetization of the spin-$1$ isotropic
Heisenberg model for several values of $d$. The dashed lines typify the discontinuity of
magnetization at the temperature of first-order phase transition.}
\vspace*{-0.27cm}
\end{figure}                       
On the other hand, further reducing of parameter $d$ induces the occurence of
discontinuity in the magnetization at the critical temperature $t_c$  what is
characteristic for the first-order phase transitions. It is noteworthy that this
discontinuity increases when the single-ion anisotropy $d$ further retreats from the value
$d_t$.  
  
\section{CONCLUSIONS}
\vspace{-0.2cm}
In this paper, the phase diagram of the spin-1 anisotropic Heisenberg model with the
uniaxial single-ion anisotropy is examined within the framework of the Oguchi's pair
approximation. We have demonstrated that both the exchange anisotropy $\Delta$ as well as
single-ion anisotropy $d$ have a significant influence on the criticality and tricriticality of
the model under investigation. Finally, it should be mentioned that the more comprehensive results of our investigation will
be published in the near future.

\vspace{0.5cm}
\noindent ACKNOWLEDGMENT: This work was supported by the grant VVGS 011/2006.\\

\end{document}